\documentclass[final]{svjour3}
\usepackage{graphicx}
\usepackage{rotating}
\usepackage{amssymb}
\usepackage{mathptmx}
\usepackage[numbers]{natbib}
\makeatletter
\journalname{Journal of Low Temperature Physics}

\usepackage{subfigure}

\newcommand{\figref}[1]{Fig.~\ref{#1}}

\begin{document}

\newcommand{\hdblarrow}{H\makebox[0.9ex][l]{$\downdownarrows$}-}
\title{Microwave Multiplexing on the Keck Array}

\author{A. Cukierman$^{1,2}$ \and Z.~Ahmed$^{1,16}$ \and S.~Henderson$^{1,16}$ \and E.~Young$^{1,2}$ \and C.~Yu$^2$ 
\and D.~Barkats$^3$ \and D.~Brown$^{16}$ \and S.~Chaudhuri$^2$ \and J.~Cornelison$^3$ \and J.~M.~D'Ewart$^{16}$ \and M.~Dierickx$^3$ \and B.~J.~Dober$^{13}$ \and J.~Dusatko$^{16}$ \and S.~Fatigoni$^5$  \and J.~P.~Filippini$^{11,12}$ \and J.~C.~Frisch$^{16}$ \and G.~Haller$^{16}$ \and M.~Halpern$^5$  \and G.~C.~Hilton$^{13}$ \and J.~Hubmayr$^{13}$ \and K.~D.~Irwin$^{1,2}$ \and K.~S.~Karkare$^{3,14}$ \and E.~Karpel$^2$ \and S.~A.~Kernasovskiy$^2$  \and J.~M.~Kovac$^3$ \and A.~Kovacs$^3$ \and S.~E.~Kuenstner$^2$ \and C.~L.~Kuo$^2$ \and D.~Li$^{16}$ \and J.~A.~B.~Mates$^{13}$ \and S.~Smith$^{16}$ \and T.~St.~Germaine$^3$ \and J.~N.~Ullom$^{13}$ \and L.~R.~Vale$^{13}$ \and D.~D.~Van~Winkle$^{16}$ \and J.~Vasquez$^{16}$ \and J.~Willmert$^9$ \and L.~Zeng$^3$ 
\and P.~A.~R.~Ade$^4$  \and M.~Amiri$^5$ \and R.~Basu Thakur$^6$ \and C.~A.~Bischoff$^7$ \and J.~J.~Bock$^{6,8}$ \and H.~Boenish$^3$ \and E.~Bullock$^9$ \and V.~Buza$^3$ \and J.~Cheshire$^9$ \and J.~Connors$^3$ \and M.~Crumrine$^9$ \and L.~Duband$^{10}$ \and G.~Hall$^{3,9}$ \and S.~Harrison$^3$ \and S.~R.~Hildebrandt$^8$ \and H.~Hui$^6$ \and J.~Kang$^2$ \and S.~Kefeli$^6$ \and K.~Lau$^9$ \and K.~G.~Megerian$^8$ \and L.~Moncelsi$^6$ \and T.~Namikawa$^{17}$ \and H.~T.~Nguyen$^{6,8}$ \and R.~O'Brient$^{6,8}$ \and S.~Palladino$^7$ \and C.~Pryke$^9$ \and B.~Racine$^3$ \and C.~D.~Reintsema$^{13}$ \and S.~Richter$^3$ \and A.~Schillaci$^6$ \and R.~Schwarz$^9$ \and C.~D.~Sheehy$^{15}$ \and A.~Soliman$^6$ \and B.~Steinbach$^6$ \and R.~V.~Sudiwala$^4$ \and K.~L.~Thompson$^2$ \and C.~Tucker$^4$ \and A.~D.~Turner$^8$ \and C.~Umilt\`{a}$^7$ \and A.~G.~Vieregg$^{14}$ \and A.~Wandui$^6$ \and A.~C.~Weber$^8$ \and D.~V.~Wiebe$^5$ \and W.~L.~K.~Wu$^{14}$ \and H.~Yang$^2$ \and K.~W.~Yoon$^2$ \and C.~Zhang$^6$}

\authorrunning{A.~Cukierman et al.}

\institute{
\email{ajcukier@stanford.edu}
\\$^1$Kavli Institute for Particle Astrophysics and Cosmology, SLAC National Accelerator Laboratory, Menlo Park, CA 94025, USA
\\$^2$Department of Physics, Stanford University, Stanford, CA 94305, USA
\\$^3$Harvard-Smithsonian Center for Astrophysics, Cambridge, MA 02138, USA
\\$^4$School of Physics and Astronomy, Cardiff University, Cardiff, CF24 3AA, UK
\\$^5$Department of Physics and Astronomy, University of British Columbia, Vancouver, BC V6T 1Z1, Canada
\\$^6$Department of Physics, California Institute of Technology, Pasadena, CA 91125, USA
\\$^7$Department of Physics, University of Cincinnati, Cincinnati, OH 45221, USA
\\$^8$Jet Propulsion Laboratory, Pasadena, CA 91109, USA
\\$^9$Minnesota Institute for Astrophysics, University of Minnesota, Minneapolis, MN 55455, USA
\\$^{10}$Service des Basses Temp\'{e}ratures, Commissariat \`{a} l'Energie Atomique, 38054 Grenoble, France
\\$^{11}$Department of Physics, University of Illinois at Urbana-Champaign, Urbana, IL 61801, USA
\\$^{12}$Department of Astronomy, University of Illinois at Urbana-Champaign, Urbana, IL 61801, USA
\\$^{13}$National Institute of Standards and Technology, Boulder, CO 80305, USA
\\$^{14}$Kavli Institute for Cosmological Physics, University of Chicago, Chicago, IL 60637, USA
\\$^{15}$Physics Department, Brookhaven National Laboratory, Upton, NY 11973, USA
\\$^{16}$SLAC National Accelerator Laboratory, Menlo Park, CA 94025, USA
\\$^{17}$Department of Applied Mathematics and Theoretical Physics, University of Cambridge, Cambridge, CB3 0WA, UK}

\maketitle

\begin{abstract}

We describe an on-sky demonstration of a microwave-multiplexing readout system in one of the receivers of the Keck Array, a polarimetry experiment observing the cosmic microwave background at the South Pole. During the austral summer of 2018-2019, we replaced the time-division multiplexing readout system with microwave-multiplexing components including superconducting microwave resonators coupled to radio-frequency superconducting quantum interference devices at the sub-Kelvin focal plane, coaxial-cable plumbing and amplification between room temperature and the cold stages, and a SLAC Microresonator Radio Frequency system for the warm electronics. In the range 5-6~GHz, a single coaxial cable reads out 528~channels. The readout system is coupled to transition-edge sensors, which are in turn coupled to 150-GHz slot-dipole phased-array antennas. Observations began in April 2019, and we report here on an initial characterization of the system performance.


\end{abstract}

\section{Introduction}

With the increase in detector count in cosmic microwave background (CMB) experiments, approaching $5 \times 10^5$ for CMB Stage~4, there is growing motivation for large readout multiplexing factors as a means of simplifying experimental designs and lowering thermal loads to low-temperature stages~\cite{cmbs4 tech}. The \emph{Keck Array} has been operating with time-division multiplexing (TDM) readout, which has been demonstrated in CMB experiments with multiplexing factors up to $\sim 100$~\cite{henderson ACT readout}. An improvement of more than an order of magnitude can be achieved with a microwave-multiplexing ($\mu$mux) readout system~\cite{irwin_lehnert, dober}. To prepare for the next generation of CMB experiments and advance the development of $\mu$mux, the TDM readout system of a single Keck receiver was replaced with a $\mu$mux system in the austral summer of 2018-2019. 

\section{Microwave Multiplexing and SMuRF Electronics}

Microwave multiplexing is a readout technology for DC-biased transition-edge sensors (TESs), which has the potential to support multiplexing factors up to~2000 for CMB experiments.
\begin{figure}[htbp]
\begin{center}
\includegraphics[width = 0.8\textwidth, keepaspectratio]{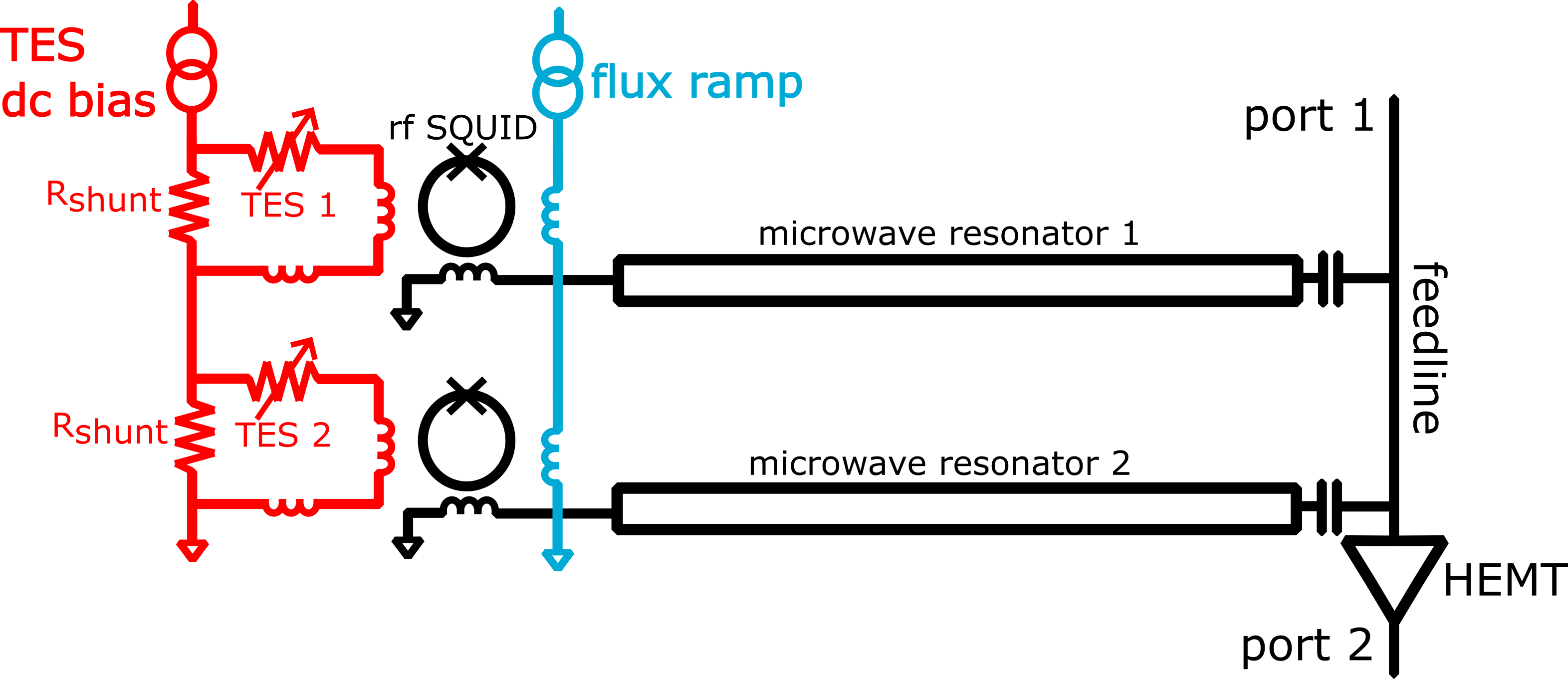}
\caption{Circuit diagram for microwave multiplexing. At port~1, probe tones are generated and sent along the feedline. Some power is shunted by $\lambda/4$ resonators, so the tones measured at port~2 are attenuated and phase shifted. Changes in the resonances produce changes in the output tones. Each resonator is in series with a small inductor coupled to an rf~SQUID, which is also coupled to a TES and the flux ramp. Current flowing in either the TES or the flux ramp shifts the resonance, which alters the output tones at port~2. \label{fig:umux_circuit} }
\end{center}
\end{figure}
A schematic of the circuit is shown in \figref{fig:umux_circuit}. A cascade of resonators with unique resonant frequencies in the range of $4$-$8~\mathrm{GHz}$ are coupled through radio-frequency (rf) superconducting quantum interference devices (SQUIDs) to TESs. The magnetic flux in the rf~SQUID controls its inductance, which in turn modulates the resonant frequency. As the TES current changes, the magnetic flux and, therefore, the resonant frequency change, i.e., the TES signal is transduced to a shift in resonant frequency. An additional inductive coupling is introduced to modulate the resonant frequency at a known rate. Typically a sawtooth current ramp, this flux bias modulates the TES signal to higher frequencies, typically $10$-$20~\mathrm{kHz}$ for CMB applications, where the $1/f$~component of the two-level-system noise in the resonators is negligibly small~\cite{mates}. The flux-ramping scheme moves the TES signal into the \emph{phase} of the fast modulation. The demodulated phase is \emph{linearly} related to the TES current, even though the SQUID response to magnetic flux is quasi-sinusoidal. For CMB applications, the width of each resonance is $\sim 100~\mathrm{kHz}$ with a depth of $10$-$20~\mathrm{dB}$. The flux ramp oscillates the resonance by $\sim 100~\mathrm{kHz}$ around the zero-current resonant frequency. 

An advantage of $\mu$mux is that the optimization of the TES bias circuit is largely independent of the readout circuit, the two being coupled only through the rf~SQUID. In particular, $\mu$mux is compatible with the DC~bias circuits of established TDM systems like those of the Atacama Cosmology Telescope (ACT) and BICEP/\emph{Keck Array}~\cite{act_instrument, BK_instrument}. A low-drift commercial voltage reference is used to provide a stable DC voltage bias.

\begin{figure}[htbp]
\begin{center}
\subfigure{\includegraphics[height=0.35\linewidth, keepaspectratio]{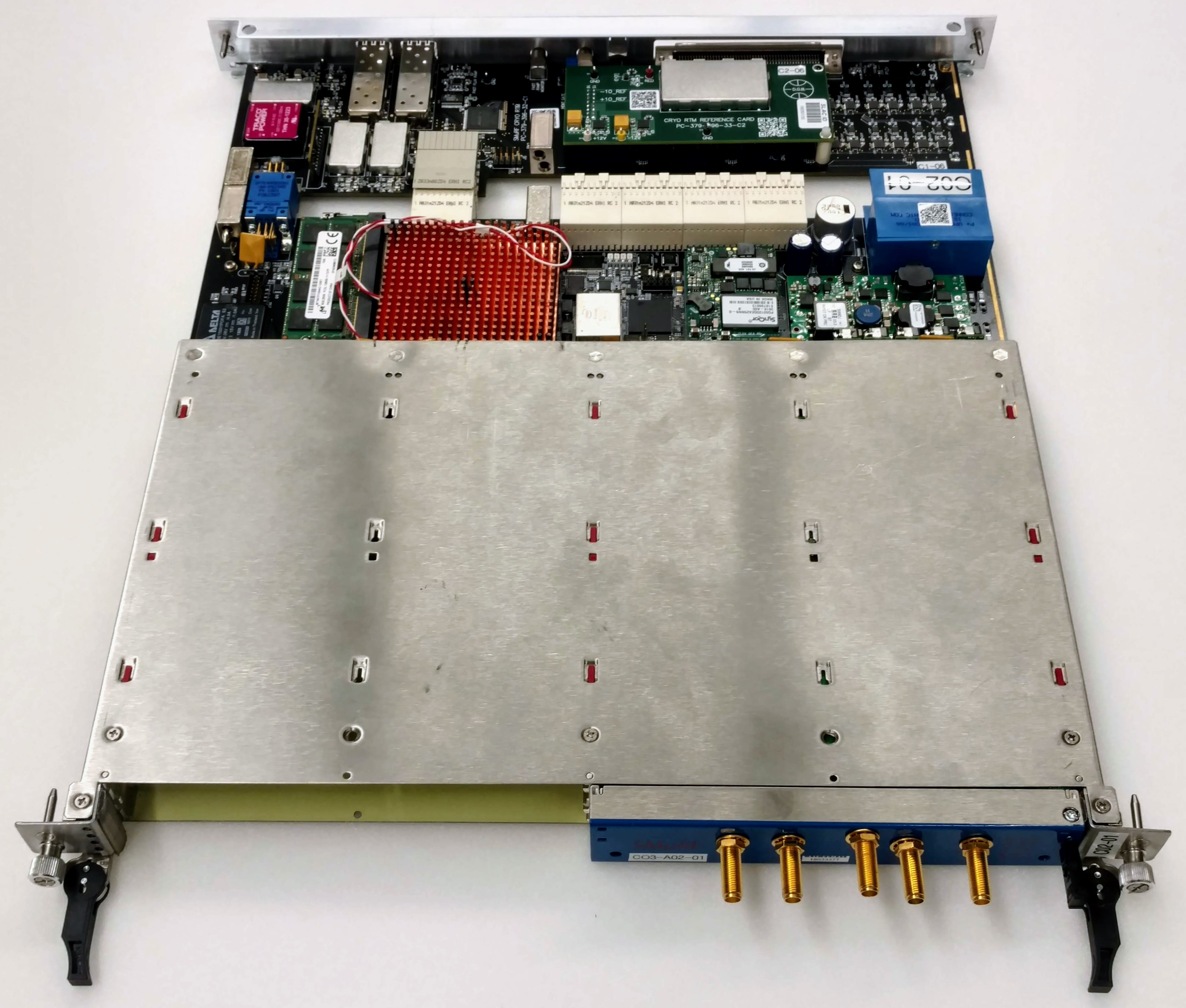}}
\hspace{0.01\textwidth}
\subfigure{\includegraphics[height=0.35\linewidth, keepaspectratio]{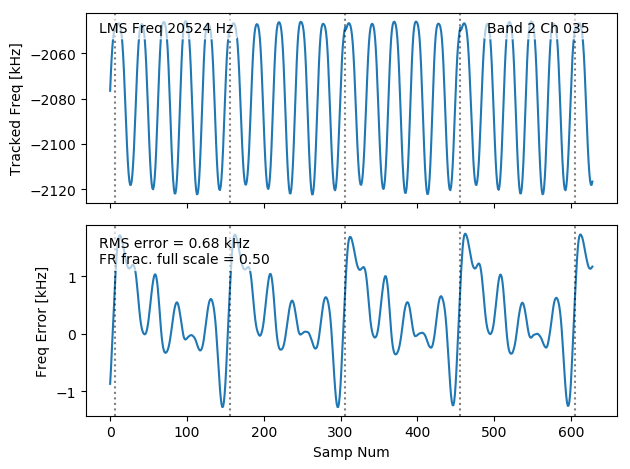}}
\caption{(\emph{Left})~FPGA carrier card with one AMC attached at bottom right and an RTM connected at top. The microwave tones travel on SMA cables connected to the AMC. The empty slot at bottom left can be used for a second AMC to double the total bandwidth from~$2$ to~$4~\mathrm{GHz}$. (\emph{Right})~The tone-tracking algorithm responding to flux-ramp modulation. The top panel shows the tracked resonant frequency, and the bottom panel shows the estimated error in the reconstruction. The flux ramp is a sawtooth current that sweeps through multiple flux quanta in each period. At the sawtooth resets, indicated by vertical dotted lines, there is a small discontinuity in the reconstruction and a larger error. The sawtooth frequency is~$4~\mathrm{kHz}$. Sweeping through five flux quanta, however, the modulation frequency is~$20~\mathrm{kHz}$. TES current alters the \emph{phase} of the oscillation. \label{fig:smurf} }
\end{center}
\end{figure}

The SLAC Microresonator Radio Frequency (SMuRF) electronics, developed at SLAC National Accelerator Laboratory, generate probe tones as well as flux-ramp and TES-bias current; after passing through the comb of resonators, the tones are returned to SMuRF, and an FPGA demodulates the TES signal from the flux-ramp oscillations~\cite{henderson_spie}. The modular system is based on SLAC's Advanced Telecommunications Computing Architecture (ATCA) Common Platform. The FPGA carrier card interfaces with Advanced Mezzanine Cards (AMCs), which generate and receive microwave tones, and a rear transition module (RTM), which provides flux-ramp and TES-bias current as well as amplifier bias voltages. At left in \figref{fig:smurf}, we show these cards integrated. 
The RTM connects to a \emph{cryostat card}, which contains bias resistors and filters and plugs directly into the cryostat. One set of cards can process 3300 tones over $4$-$8~\mathrm{GHz}$. 

In SMuRF, the excitation tones are updated in real time to follow the transmission minima of the true resonators. This \emph{tone tracking} allows SMuRF to lower the input power to the HEMT amplifier and improve linearity in the transmission. As the resonances move in response to TES current or flux ramp, a fast feedback loop shifts the probe tones to remain always on resonance. Tone tracking increases the number of tones that may be operated simultaneously before nonlinearities in the HEMT create intermodulation distortion products that can act as an effective noise floor~\cite{henderson_spie}. At right in \figref{fig:smurf}, we show the tone-tracking algorithm responding to flux-ramp modulation. 

A Python-based user control software called {\tt pysmurf} interacts with the SMuRF cards through an Experimental Physics and Industrial Control System (EPICS) server. For users as well as for higher-level telescope control systems, {\tt pysmurf} allows control of amplifiers, microwave-tone generation and readback, flux-ramp modulation, tone tracking and TES bias current. The software also controls data streaming and provides analysis functions for noise and TES parameter estimation.

\section{Retrofitting the Keck Array}

The \emph{Keck Array}, part of the BICEP/Keck program, is a CMB polarimetry experiment operating at the South Pole since 2012~\cite{BK15}. The project targets degree-scale $B$-mode polarization to detect or constrain the presence of primordial gravitational waves~\cite{seljak_zaldarriaga}. The \emph{Keck Array} consists of five receivers, each with an aperture of~$26~\mathrm{cm}$ and sensitive to millimeter wavelengths over a $\sim 30\%$ bandwidth. The receivers are installed on the same mount and scan the sky together. Each receiver contains $528$~TESs cooled to $250~\mathrm{mK}$. Optical power is received by slot-dipole phased-array antennas and routed by microstrip lines to the TESs~\cite{BK_instrument}. All receivers have used TDM readout until the 2019 season.
 
To understand integration challenges and improve technological readiness of $\mu$mux and SMuRF, a single Keck receiver was retrofitted with $\mu$mux readout on a previously deployed $150$-$\mathrm{GHz}$ focal-plane unit (FPU). Observations began in early 2019. This retrofit was implemented as a mission of opportunity before the decommissioning of the \emph{Keck Array} in the austral summer 2019-2020, when the \emph{BICEP Array} will take its place~\cite{BA}.

Two types of readout chips were fabricated at NIST, Boulder. The \emph{resonator chips} contain the resonators and rf~SQUIDs. The feedline, realized as a coplanar waveguide, and flux-ramp lines run along them; multiple chips are strung together by bonding the feedlines and flux-ramp leads to each other. Each resonator chip covers a unique $125$-$\mathrm{MHz}$ bandwidth. The resonances are designed to be separated by $2~\mathrm{MHz}$. We only used the range $5$-$6~\mathrm{GHz}$, but the resonator density in frequency space is appropriate for a multiplexing factor of~2000 if the full SMuRF bandwidth of $4$-$8~\mathrm{GHz}$ were utilized. The multiplexing factor is limited by requirements on the resonators and not by SMuRF. The \emph{shunt chips} contain the TES bias resistors and bandwidth-limiting inductors. Each resonator chip is paired with a single shunt chip, though all the shunt chips are identical and interchangeable. Each resonator/shunt chip pair biases and reads out 64 TESs on the same bias circuit.

\begin{figure}[htbp]
\begin{center}
\subfigure{\includegraphics[height=0.3\linewidth, keepaspectratio]{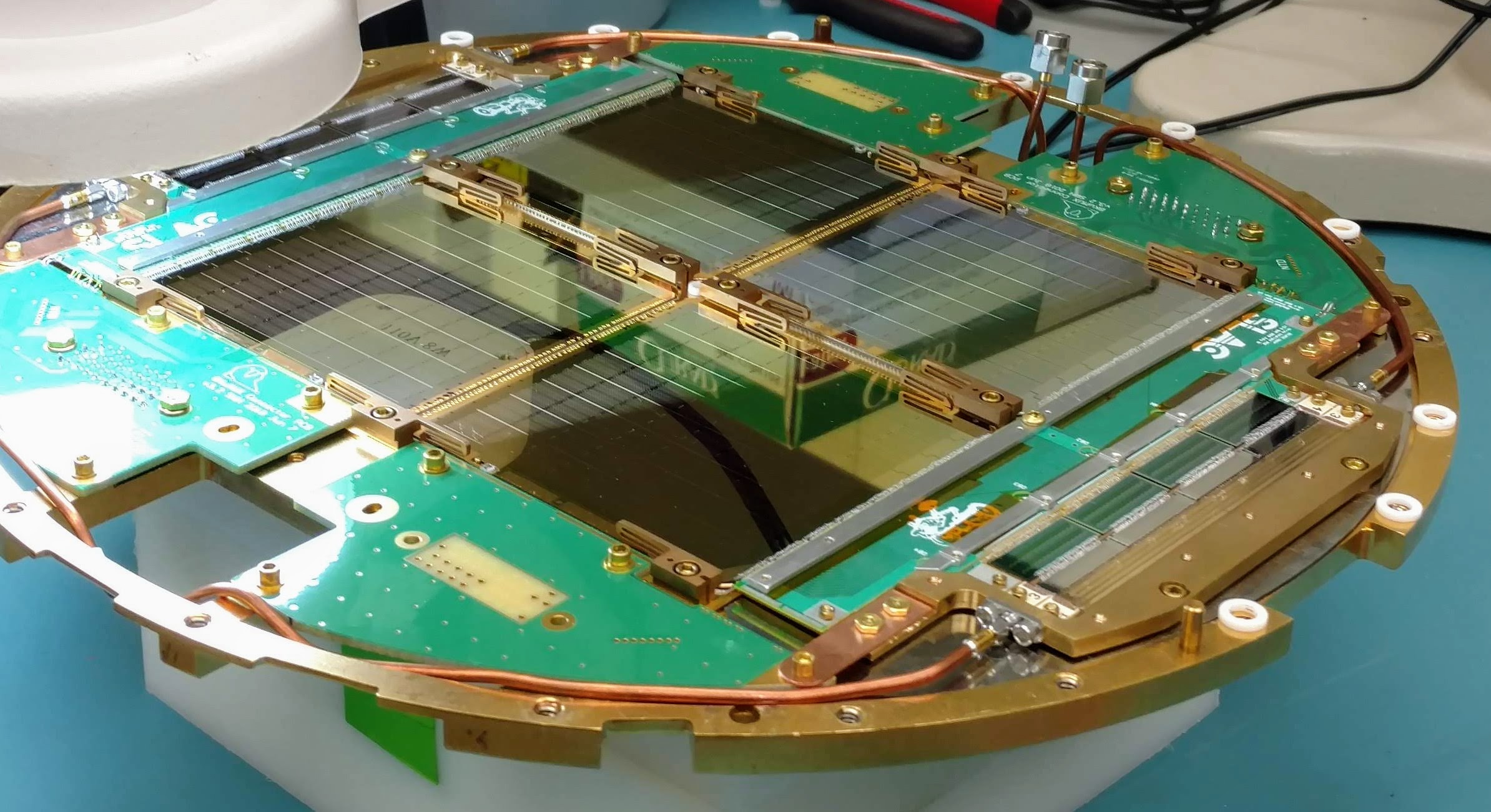}}
\hspace{0.01\textwidth}
\subfigure{\includegraphics[height=0.3\linewidth, keepaspectratio]{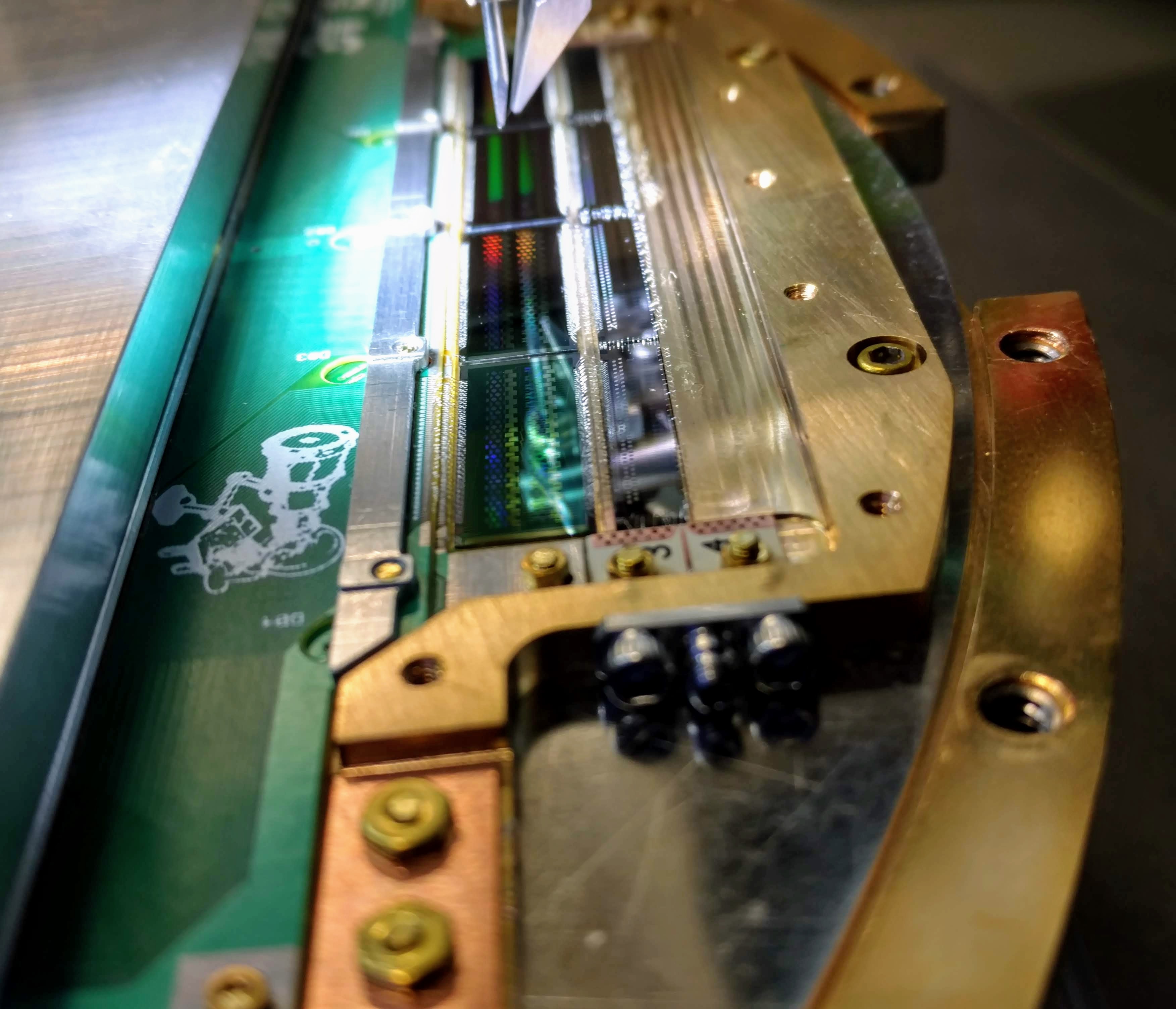}}
\caption{(\emph{Left})~The retrofitted FPU with the sky side facing down. In the center are four detector tiles, which integrate the TESs and antennas and were fabricated at the JPL Microdevices Laboratory. At lower right and upper left are the two RF modules containing the resonator and shunt chips. The TES bias and flux ramp enter from the bottom of the PCBs at left and upper right. The microwave tones enter and exit through the pair of SMA connectors at upper right. To keep a low profile, SuperMini coaxial cables from Southwest Microwave were used to route the tones around the FPU. In operation, the entire face-up side of the FPU is protected by a $\mathrm{Nb}$~cover, which acts as a $\lambda/4$ backshort for the antennas and as a magnetic shield for the SQUIDs. (\emph{Right})~A close-up view of one RF module during the wire-bonding process. The module contains four resonator chips at center right and four shunt chips at center left. The first and fourth chips are bonded to PCBs to which the SuperMini connectors are soldered. At left are PCBs with $\mathrm{Al}$~traces which bring the TES leads to the detector tiles. To achieve the necessary bond pad pitch, two such PCBs were stacked and offset. \label{fig:fpu} }
\end{center}
\end{figure}

We replaced the TDM SQUIDs on a $150$-$\mathrm{GHz}$ Keck FPU, which had been used for observations in 2013-2016, with the resonator and shunt chips described above (see \figref{fig:fpu}). 
The $\mu$mux components were divided into two RF modules, each mating with two detector tiles (256 TESs) and covering $500~\mathrm{MHz}$ of microwave bandwidth. The RF modules were designed to respect the existing $1/8"$ vertical clearance of the FPU without additional modifications to the Keck design. The RF modules were first tested without detector tiles. The detectors were divided into eight bias circuits, requiring eight pairs of leads connected to the FPU. A single pair of flux-ramp leads entered the cryostat and was split at the $4$-$\mathrm{K}$ stage to send half the current to each RF module. 

\begin{figure}[htbp]
\begin{center}
\subfigure{\includegraphics[height = 0.45\linewidth, keepaspectratio]{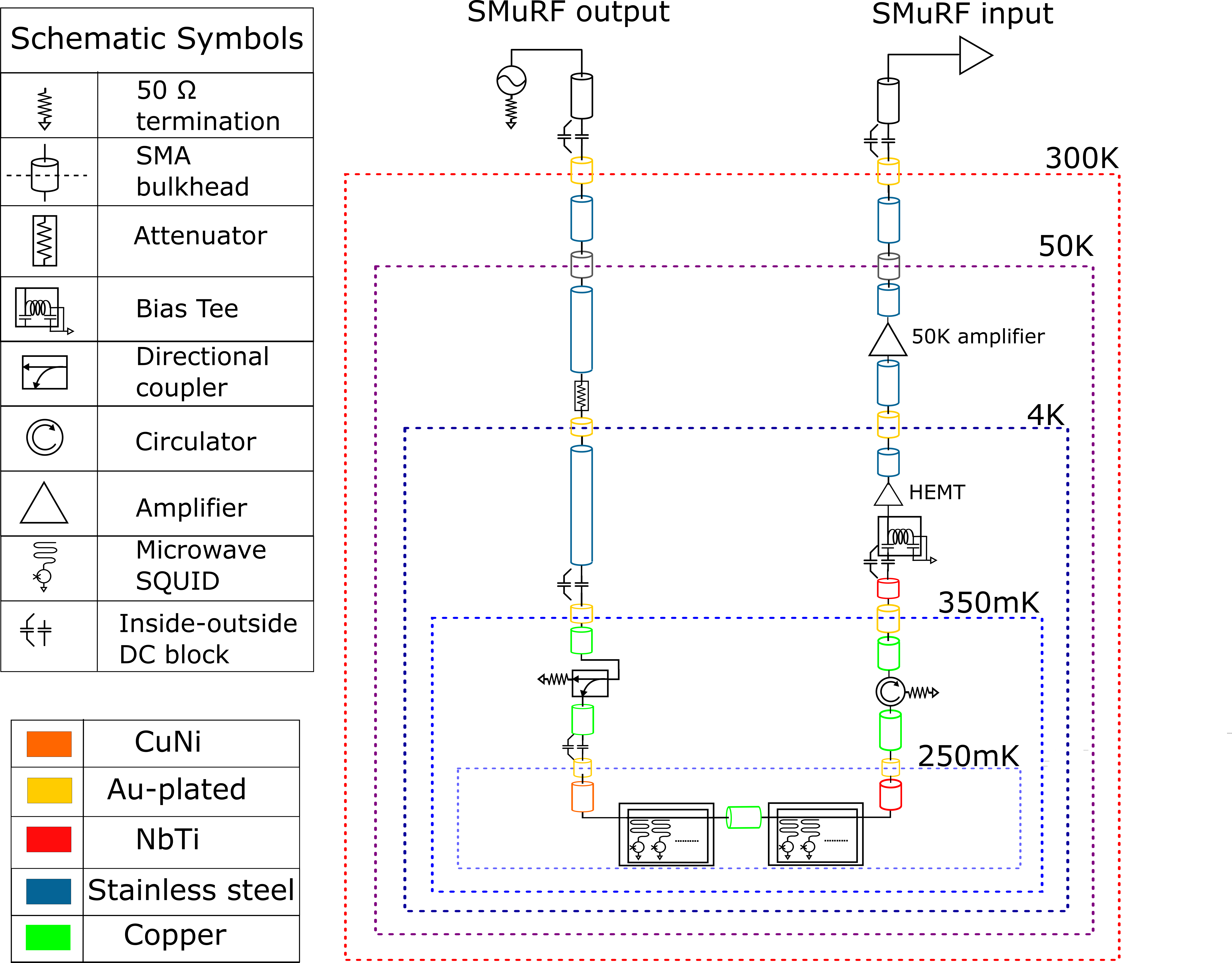}}
\hspace{0.01\textwidth}
\subfigure{\includegraphics[height = 0.45\linewidth, keepaspectratio]{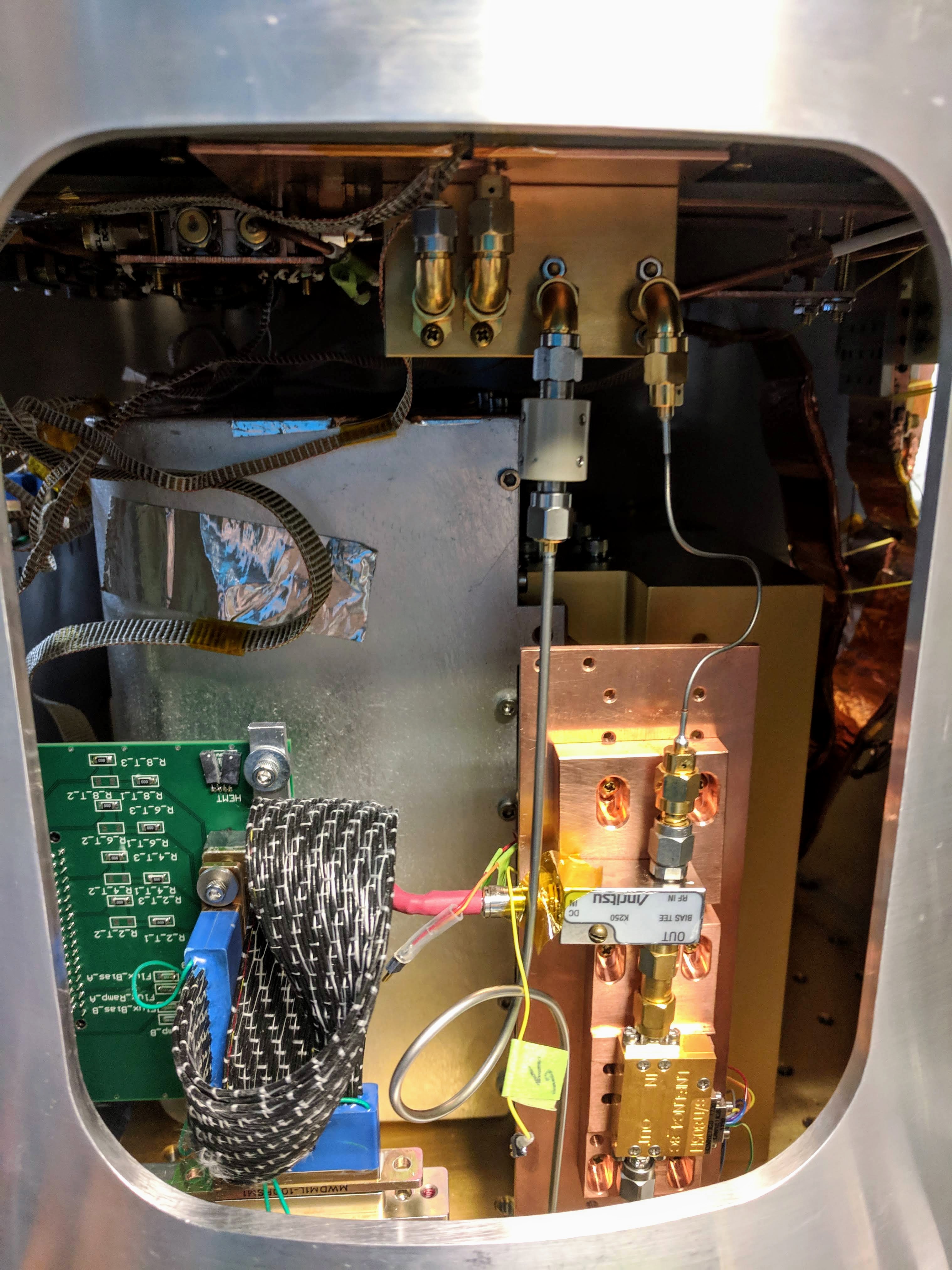}}
\caption{(\emph{Left}) Schematic of the microwave chain installed in the Keck receiver. Tones are generated at the SMuRF output port and attenuated to an appropriate level for the cold resonators. On the way back out, the tones are amplified by a Low Noise Factory HEMT at~$4~\mathrm{K}$ and by a B\&Z Technologies amplifier at~$50~\mathrm{K}$. Other components and materials are chosen to suppress reflections, heat sink the coaxial cables and isolate the temperature stages. \emph{(Right)} The $4$-$\mathrm{K}$ stage showing the HEMT and Anritsu bias tee. \label{fig:rfchain} }
\end{center}
\end{figure}

A microwave chain was installed in the receiver (see \figref{fig:rfchain}). 
Probe tones output by SMuRF enter the cryostat through a single SMA connection. To avoid saturating the rf~SQUIDs, the tones are attenuated as they travel to the FPU. We include DC~blocks and stainless-steel coaxial cables to isolate temperature stages, and we use a circulator to suppress reflections. A directional coupler presents a cold resistive load to the resonators. 

The microwave tones enter the FPU through a single SMA connection, pass through both RF modules in series and then exit the FPU through a single SMA connection. On the way back to room temperature, the gain is increased by $\sim 20~\mathrm{dB}$ by a HEMT at~$4~\mathrm{K}$ and again by $\sim 10~\mathrm{dB}$ by an amplifier at~$50~\mathrm{K}$. The double-amplifier design improves linearity of the RF chain and allows more tones to be operated simultaneously~\cite{henderson_spie}. A bias tee heat sinks the HEMT to~$4~\mathrm{K}$. The tones exit the cryostat through a single SMA connection and return to the AMC. A room-temperature amplifier boosts the gain by an additional~$20~\mathrm{dB}$.
Since a Keck receiver contains only $528$~TESs, only $1$-$\mathrm{GHz}$ of microwave bandwidth was necessary. We, therefore, ran the system with a single AMC producing tones in the range of $5$-$6~\mathrm{GHz}$.

The RF chain was qualified at Stanford University, and the RF modules were tested \emph{without} detector tiles. The detector tiles were then integrated into the FPU and wire bonded to the RF modules. The SMuRF electronics were integrated with the Generic Control Program ({\tt GCP}) that controls the \emph{Keck Array} and handles data acquisition (DAQ). At the Center for Astrophysics (CfA) at Harvard University, a receiver identical to those on the \emph{Keck Array} was used to test the retrofit in both dark and optical configurations.

\section{Deployment to the South Pole}

\begin{figure}[htbp]
\begin{center}
\subfigure{\includegraphics[height =0.26\textwidth, keepaspectratio]{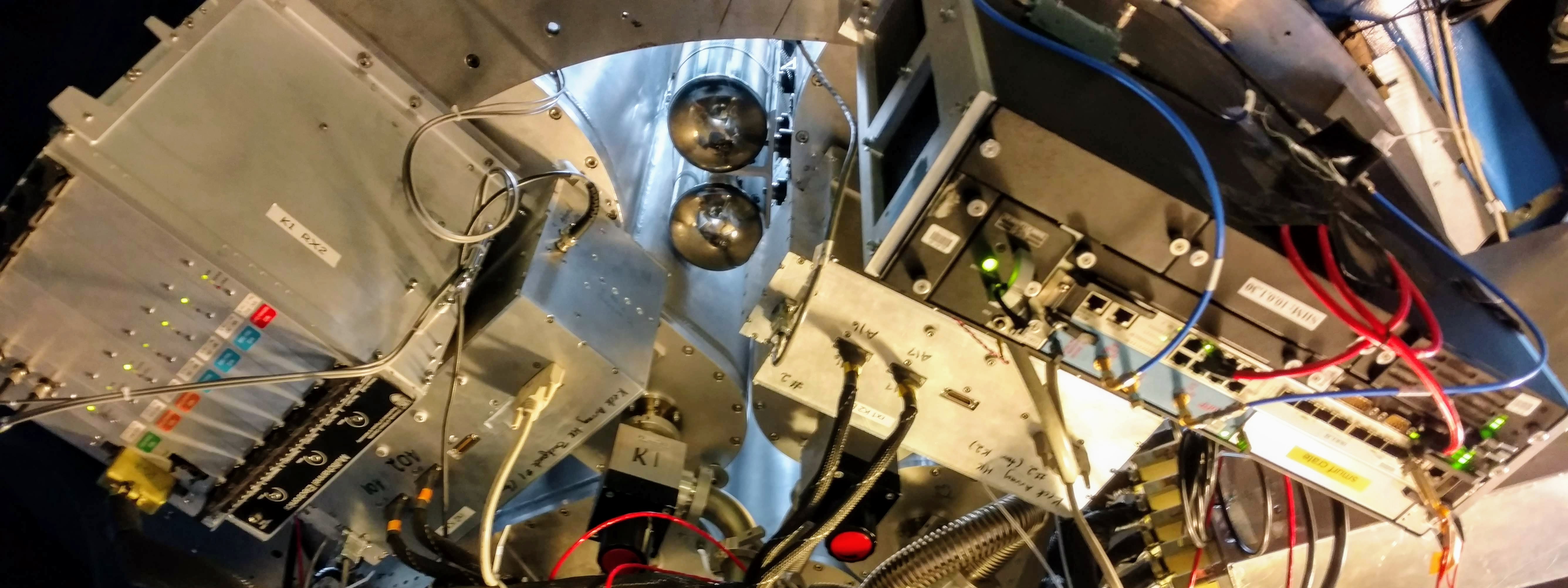}}
\hspace{0.01\textwidth}
\subfigure{\includegraphics[height = 0.26\textwidth, keepaspectratio]{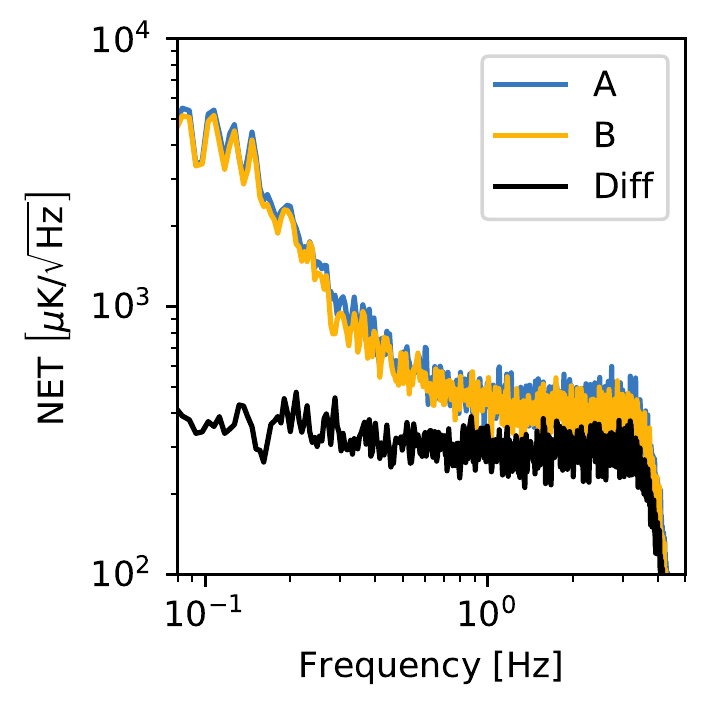}}
\caption{(\emph{Left}) View from inside the Keck telescope mount. At left is an unaltered Keck receiver with an MCE crate. At right is a SMuRF crate attached to the $\mu$mux receiver. Data and timing signals are passed through fiber-optic cables that are routed down the telescope mount to stationary servers below. (\emph{Right}) Sample PSDs from a period of relatively good atmospheric conditions. The PSD is parameterized as noise-equivalent temperature (NET) fluctuations of the CMB. The blue and yellow lines show PSDs after third-order polynomial filtering for a detector pair. The black line shows the PSD of the uncorrelated component, the \emph{pair difference}. The large rise in noise at low frequencies in the individual detectors, mostly due to unpolarized atmospheric fluctuations, is greatly suppressed in the pair difference, where the polarization signal lies. The roll-off above $5~\mathrm{Hz}$ is an artifact of the DAQ down-sampling and filtering. For Keck, the science band is $0.1$-$1~\mathrm{Hz}$.  \label{fig:crate} }
\end{center}
\end{figure}

In late~2018, the $\mu$mux components from the Harvard tests were shipped to the South Pole. One of the $220$-$\mathrm{GHz}$ receivers was pulled from the telescope mount and retrofitted with the $150$-$\mathrm{GHz}$ $\mu$mux FPU. The optics tube was replaced to match the new frequency band, and the RF chain was installed. Testing was performed in a lab environment, and the SMuRF electronics was integrated with the Keck timing, control and DAQ systems. Finally, the receiver was reinstalled in the telescope mount. The crate of SMuRF cards was installed directly on the Keck receiver (see \figref{fig:crate}) and replaces the Multi-Channel Electronics (MCE) that was used to read out that receiver in previous years.
A SMuRF control server intercepts commands from the main Keck control server, translates to the {\tt pysmurf} equivalents and then passes the commands on to the SMuRF crate. 

There were several sources of yield hits. Some resonator chips did not yield in fabrication in time for polar deployment. Some resonators collided with each other in frequency space, making them impossible to track. An unfortunate but fixable design of the TES bias circuit created a vulnerability to inter-TES electrical shorts that prevented bias current from reaching many TESs, but this was understood too late to be mended. There are also some yield hits from the detector tiles themselves. Altogether, we see 120 optically active TESs.

\begin{figure}[htbp]
\begin{center}
\subfigure{\includegraphics[height=0.24\linewidth, keepaspectratio]{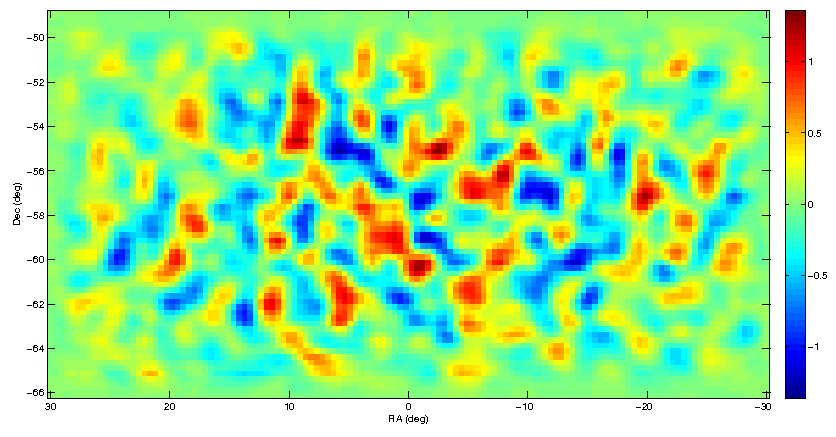}}
\hspace{0.01\textwidth}
\subfigure{\includegraphics[height=0.24\linewidth, keepaspectratio]{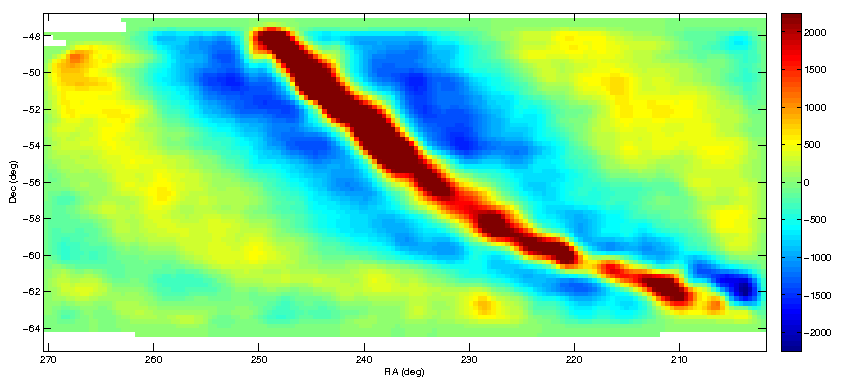}}
\caption{Preliminary temperature maps at $150~\mathrm{GHz}$ from the Keck receiver retrofitted with microwave-multiplexing readout. The $x$-axis indicates right ascension (RA), and the $y$-axis indicates declination. These maps were made from approximately four weeks of observations with approximately 60 of the optically active TESs. (\emph{Left}) CMB temperature anisotropies. (\emph{Right}) A section of the Milky Way galaxy.  \label{fig:maps} }
\end{center}
\end{figure}

The $\mu$mux data streaming to the Keck DAQ system looks identical to the data collected by the MCE readout and can be integrated into the established BICEP/Keck analysis pipeline. The mapping from resonant frequency to physical detector was validated through far-field beam mapping before the 2019 observing season began. Each antenna is connected to a pair of TESs, one for each linear polarization. The polarized sky signal is estimated from the difference between the responses of the paired TESs. Sample power spectral densities (PSDs) are shown at right in \figref{fig:crate} for one of the best detector pairs. We see a range of noise performance, many pair-differenced PSDs showing a rise at low frequencies proportional to the TES responsivities and possibly due to vibrational or thermal sensitivity of the FPU. This behavior is not seen in laboratory tests in other $\mu$mux systems. Overbiased or superconducting, the performance at the South Pole is broadly consistent with Johnson noise, which indicates that the spurious noise is entering through the TESs rather than the readout circuit. Preliminary maps of sky temperature from approximately four weeks of observations are shown in \figref{fig:maps}.
System characterization will continue through the end of the 2019 season. 

\section{Future Directions}

We demonstrated an end-to-end integrated microwave-multiplexing readout system for use in CMB experiments. 
The \emph{Simons Observatory}, a CMB polarimetry experiment that will begin observing from the Atacama Desert in 2020, will use $\mu$mux and SMuRF to read out $\sim 60,000$ TESs~\cite{SO}.
The \emph{BICEP Array} is considering $\mu$mux for its highest-frequency receiver, which will contain more than 20,000 TESs operating simultaneously at $220$ and $270~\mathrm{GHz}$~\cite{BA}.
Beyond CMB applications, SMuRF and $\mu$mux are being developed for use with X-ray TES calorimeters for both beamline and astrophysical satellite experiments. Non-TES detectors such as metallic magnetic calorimeters (MMCs) can also be coupled to a $\mu$mux readout architecture~\cite{mmc}, and the SMuRF electronics can be used for resonator-based detectors such as kinetic-inductance detectors (KIDs).

\begin{acknowledgements}
This work was supported by the U~S Department of Energy, Office of Science, Contract DE-AC02-76SF00515. CY was supported by the NSF GRFP.
\end{acknowledgements}

\pagebreak

\end{document}